\documentclass[aps,prb,twocolumn,superscriptaddress,showpacs,fleqn,floatfix]{revtex4}

\usepackage[dvips]{epsfig}
\usepackage[dvips]{color}
\usepackage{amssymb}

\newcommand{\zncr}{ZnCr$_2$O$_4$}
\newcommand{\figwidth}{0.8\columnwidth}

\newcommand{\lvec}[1]{\mathbf{l}^{(#1)}}
\newcommand{\lij}[2]{l_{#1}^{(#2)}}
\newcommand{\lx}[1]{l_x^{(#1)}}
\newcommand{\ly}[1]{l_y^{(#1)}}
\newcommand{\lz}[1]{l_z^{(#1)}}

\begin{document}
\title{Evidence for orthorhombic distortions in the ordered state of ZnCr$_2$O$_4$: A magnetic resonance study }
\author{V.N. Glazkov}
\affiliation{P. L. Kapitza Institute for Physical Problems RAS,
119334 Moscow, Russia}

\author{A.M. Farutin}
\affiliation{P. L. Kapitza Institute for Physical Problems RAS,
119334 Moscow, Russia}

\author{V. Tsurkan}
\affiliation{Experimental Physics V, Center for Electronic
Correlations and Magnetism, University of Augsburg, 86135 Augsburg,
Germany}

\affiliation{Institute of Applied Physics, Academy of Sciences of
Moldova, MD-2028 Chi\c{s}in\v{a}u, R. Moldova}

\author{H.-A. Krug von Nidda}
\affiliation{Experimental Physics V, Center for Electronic
Correlations and Magnetism, University of Augsburg, 86135 Augsburg,
Germany}

\author{A. Loidl}
\affiliation{Experimental Physics V, Center for Electronic
Correlations and Magnetism, University of Augsburg, 86135 Augsburg,
Germany}
\date{\today}

\begin{abstract}
We present an elaborate electron-spin resonance study of the
low-energy dynamics and magnetization in the ordered phase of the
magnetically frustrated spinel \zncr{}. We observe several resonance modes
corresponding to different structural domains and found that the
number of domains can be easily reduced by field-cooling the sample
through the transition point.  To describe the observed
antiferromagnetic resonance spectra it is necessary to take into
account an orthorhombic lattice distortion in addition to the
earlier reported tetragonal distortion which both appear at the
antiferromagnetic phase transition.
\end{abstract}

\maketitle

\section{Introduction}

The intriguing physics  of  spinel compounds is in the focus of
current solid-state research. The current hot debates on the origin
of exotic phenomena and ground states in magnetic spinels concern,
e.g., the Verwey transition in Fe$_3$O$_4$
\cite{huang:04,leonov:04}, heavy-fermion formation in LiV$_2$O$_4$
\cite{kondo:97,krimmel:99}, colossal magnetoresistance in Cu doped
FeCr$_2$S$_4$ \cite{ramirez:97,fritsch:03}, gigantic Kerr rotation
\cite{ogushi:05} and the orbital glass state in FeCr$_2$S$_4$
\cite{fichtl:05}, the spin-orbital liquid  in FeSc$_2$S$_4$
\cite{fritsch:04}, the colossal magneto-capacitive effect in
CdCr$_2$S$_4$ and HgCr$_2$S$_4$ \cite{hemberger:05,weber:06}, the
negative thermal expansion and strong spin-phonon coupling in
ZnCr$_2$Se$_4$ and ZnCr$_2$S$_4$ \cite{hemberger:07,rudolf:07,
hemberger:06}, the spin dimerization in CuIr$_2$S$_4$
\cite{radaelli:02} and MgTi$_2$O$_4$ \cite{schmidt:04}, and the
spin-Peierls-like transitions  in 3-dimensional solids
\cite{lee:00,tchernyshyov:02a,tchernyshyov:02b}.  The appearance of
these fascinating ground states is attributed to the competition of
charge, spin and orbital degrees of freedom, which are strongly
coupled to the lattice.

Additional complexity in the normal \textit{AB}$_2$\textit{X}$_4$ spinels arises from
the frustration effects related to the topological constrains of the
pyrochlore lattice of corner-sharing tetrahedra of the \textit{B}-site
magnetic ions. In this geometry the exchange interaction alone
cannot select a unique ground state. As a result, the magnetic
system remains in the disordered state down to temperatures much
lower than the scale provided by the exchange interaction. In
\zncr{} strong direct antiferromagnetic Cr-Cr exchange is manifested
by the Curie-Weiss temperature of about $-400$~K, while magnetic
order appears only below 12~K via a first-order phase transition. At
this transition the aforementioned degeneracy is lifted by a
structural deformation, which is reported to be
tetragonal.\cite{lee:00}

Neutron-scattering experiments have proven that  noncollinear
commensurate antiferromagnetic order is established below the
transition temperature. However, the details of the magnetic
structure are still under heavy debate. It was speculated that a
multi-\textit{k} structure is formed.\cite{chung-prl, lee-jpcm} Moreover,
sample dependent intensities of the magnetic reflections  suggest
that \zncr{} is critically located close to several spin
structures\cite{chung-prl}.

In the present paper we report results of a magnetic resonance study
in the ordered phase of \zncr{}. Earlier ESR studies
\cite{martinho-prb} concentrated on the paramagnetic state above
$T_N$ and did not report any resonance absorption signals below
$T_N$. We find several gapped resonance modes, spin-reorientation
transitions, and evidences for an orthorhombic structural
deformation. We observe magnetic resonance signals originating from
different structural domains. We demonstrate that these domains can
be effectively aligned by field cooling in a moderate magnetic
field. The low-energy dynamics can be described qualitatively within
the exchange-symmetry theory, indicating that magnetic order in
\zncr{} is governed by a single order parameter.

\section{Experimental details}

ZnCr$_2$O$_4$ single crystals were grown by chemical transport
reactions from polycrystalline starting material prepared by
solid-state reactions of stoichiometric binary zinc and chromium oxides of
99.99\% purity. Perfect single crystalline samples of octahedral
shape and dimensions up to 3 mm on the edge were obtained. X-ray
diffraction at room temperature revealed a single-phase material
with the cubic spinel structure with a lattice constant $a =
8.332(1)$~{\AA} and an oxygen fractional coordinate $x = 0.263(1)$.
The magnetic properties were studied using a commercial SQUID
magnetometer (Quantum Design MPMS-5) working at fields up to 50~kOe.

Magnetic resonance measurements in the wide frequency range from
20 to 150~GHz were performed at the Kapitza Institute. For these
measurements we have used a set of home-made transmission-type ESR
spectrometers equipped with a superconducting cryomagnet.
High-sensitivity X-band (9.3~GHz) magnetic resonance experiments
were carried out using a Bruker "Elexsys E500 CW" spectrometer
equipped with an Oxford Instruments helium gas-flow cryostat.
Magnetic resonance absorption spectra were recorded at different
frequencies for three principal orientations of the magnetic
field:
$\mathbf{H}||\langle001\rangle,\langle110\rangle,\langle111\rangle$.
The measurements were mostly done on zero-field cooled samples, the effect
of field cooling was checked at certain frequencies.

\section{Experimental results.}

\begin{figure}
  \centering
  \epsfig{file=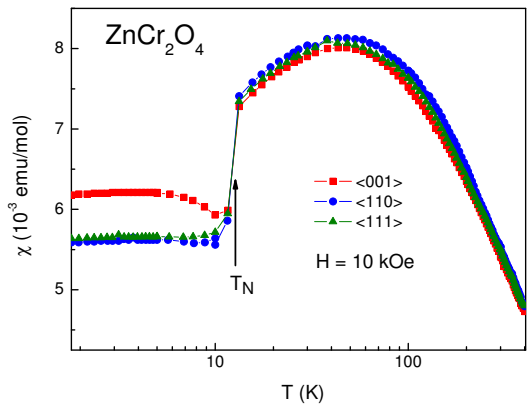, width=\figwidth, clip=}
  \caption{(color online) Temperature dependences of the magnetic susceptibility in
  different principal orientations.  All curves are measured on cooling in the field of 10 kOe.}\label{fig:chi}
\end{figure}

\begin{figure}
  \centering
  \epsfig{file=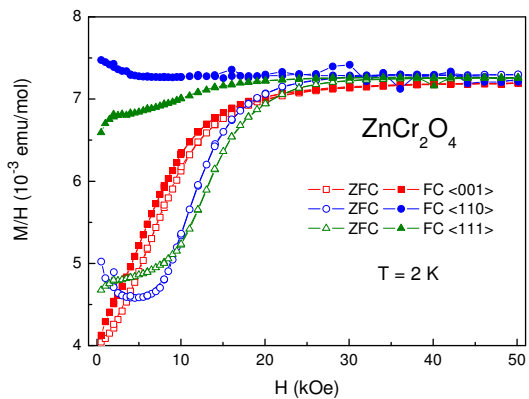, width=\figwidth, clip=}
  \caption{(color online) Field dependences of the magnetization
  divided by field  for the zero-field-cooled (ZFC, open symbols) sample and field-cooled
  (FC, closed symbols) for different orientations at T=2K.}\label{fig:m(h)}
\end{figure}

Figure \ref{fig:chi} shows the temperature dependence of the
magnetic susceptibility of \zncr{} single crystals for the
magnetic field applied along all three characteristic directions
of the cubic system. The susceptibility is isotropic above the
antiferromagnetic (AFM) transition temperature $T_N=12.5$~K. At
high temperatures the susceptibility follows a Curie-Weiss law,
but deviates already at about 100 K and develops a broad maximum
around 50 K indicative for short-range AFM correlations. At $T_N$
one observes a discontinuous change of the data typical for a
first-order transition. In the magnetically ordered regime the
magnetic susceptibility of the zero-field-cooled (ZFC) sample
shows a pronounced anisotropy with the highest value for
measurements along the $\langle001\rangle$ direction (for the
measurements in the field of 10 kOe).

To investigate the anisotropy in more detail, the magnetization
was measured dependent on the magnetic field both for zero-field
cooling (ZFC) as well as after field-cooling (FC) in 50 kOe. As
shown in Fig. \ref{fig:m(h)}, the ZFC data manifest a non-linear
behavior  in small fields (up to 20 kOe) and a linear increase of
the magnetization for higher fields. For
$\mathbf{H}||\langle100\rangle$ the M/H curve shows a linear
increase indicating smooth rotation of the magnetization, while
for the two other directions strong non-linearities are observed
around 15 kOe, as typical for a spin-flop. Field cooling has only
a weak effect for $\mathbf{H}||\langle001\rangle$, but leads to
nearly constant M/H for the other two orientations.  In the field
cooled sample the largest value of magnetic susceptibility is
observed for $\mathbf{H}||\langle110\rangle$.

\begin{figure}
  \centering
  \epsfig{file=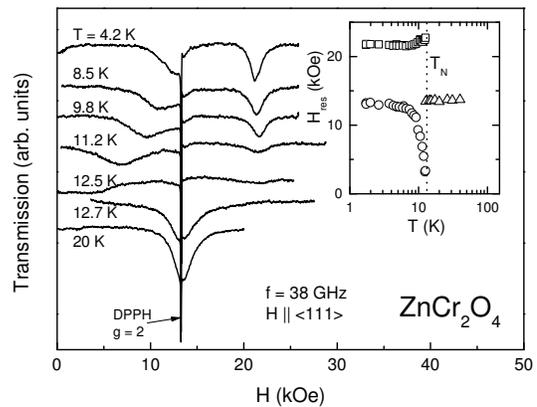, width=\figwidth, clip=}
  \caption{Field dependences of microwave absorption at different
  temperatures (ZFC sample). Inset: temperature dependence of the resonance
  fields.  The narrow line at $H=13$~kOe is a DPPH
  $(g=2.0)$ marker.}\label{fig:38ghz}
\end{figure}

The evolution of the resonance-absorption spectrum with
temperature is shown in Figure \ref{fig:38ghz}. At high
temperatures (in the paramagnetic phase) a single absorption
component with a $g$-factor close to 2.0 is observed. The
transition to the antiferromagnetically ordered state is clearly
marked by the discontinuous transformation of the  resonance
absorption spectrum: below the N\'{e}el temperature $T_N$ the
absorption spectrum consists of several components strongly
shifted from the paramagnetic resonance position. No hysteresis
exceeding the resolution limit of 0.1~K was detected at the
transition. On cooling below $T_N$ the resonance lines first show
a pronounced shift, but below 5~K the temperature dependence of
the resonance positions is negligible.

\begin{figure}
  \centering
  \epsfig{file=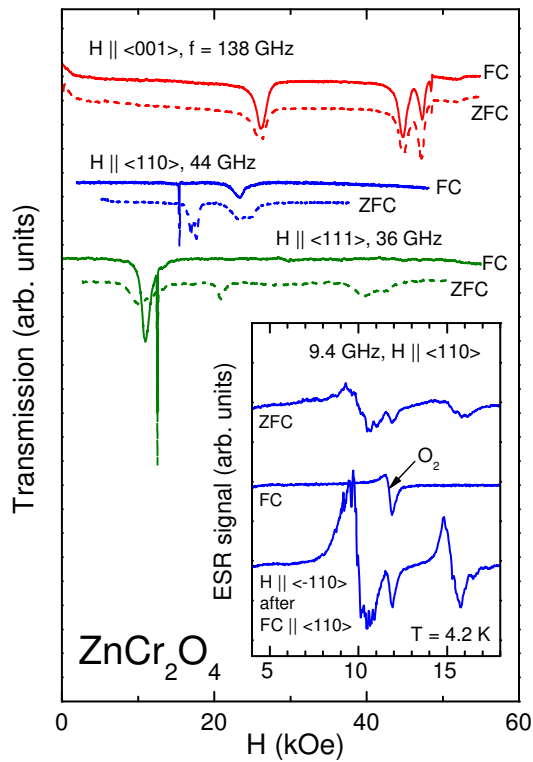, width=\figwidth, clip=}
  \caption{(Color online) Comparison of the resonance absorption observed at 1.8~K
  in zero-field cooled (ZFC) and field-cooled (FC)
  samples for different orientations of the magnetic
  field. Inset: Field cooling effect on the AFM soft mode at
  $T=4.2$~K and $f=9.4$~GHz. After FC  only the
  background signal remains due to frozen oxygen).}\label{fig:fczfcall}
\end{figure}

The shape of the resonance absorption spectra is strongly affected
by field cooling. Figure \ref{fig:fczfcall} compares the resonance
absorption measured on ZFC and FC samples. Here field cooling was
performed at the field of 50~kOe starting from 20~K. For
$\mathbf{H}||\langle110\rangle$ and $\langle111\rangle$ field
cooling leads to the disappearance of some of the absorption
components. The remaining absorption components are usually
slightly shifted from the corresponding absorption component
measured on the ZFC sample. The vanishing absorption intensity
does not necessarily add to the remaining components: for example,
for $\mathbf{H}||\langle110\rangle$ the intensity of the remaining
component after field cooling is the same as for the ZFC sample.
For $\mathbf{H}||\langle001\rangle$ field cooling effects are less
evident --- all absorption components are observed in FC samples,
field cooling leads only to a slight change of the absorption
intensity.

Actually, application of a field of 50~kOe  during the cooling
seems to be excessive. It is enough to cool the sample at the
moderate field of 18~kOe to suppress some of the resonance modes
as can be seen in the inset of Figure \ref{fig:fczfcall}.
Especially, the soft modes indicative for spin-reorientation,
disappear after field cooling but reappear as the sample is
rotated to another crystallographically equivalent position. Note
that the nonlinearity of the magnetization curves vanishes after
field cooling as well. The stability of the ZFC resonance
absorption under prolonged exposure to the magnetic field below
$T_N$ was also checked by the high-sensitive X-band measurements.
At 4~K the shape of the resonance absorption is reproducible to
the finest details. However, at 8~K (which is still below $T_N$)
keeping the $\mathbf{H}||\langle110\rangle$-oriented sample at
18~kOe for 90 minutes leads to 30\% reduction of the observed
resonance absorption.

The entire frequency-field diagrams for the different orientations
of the magnetic field are given in Figure \ref{fig:fit3}. These
dependences demonstrate the presence of several resonance modes with
zero-field gaps of $21\pm2$~GHz and $113\pm2$~GHz. For
$\mathbf{H}||\langle110\rangle$ and $\langle111\rangle$ one of the
resonance modes softens  in the magnetic field between 10 and
15~kOe. Note that the nonlinearity of the magnetization curves is
also observed at $H=10\ldots15$~kOe for these directions (Figure
\ref{fig:m(h)}). Field cooling reduces the number of the observed
resonance modes to two: one for each of the zero-field gaps.

\section{Discussion.}

\subsection{Phase transition, domains, field cooling}

As documented above, the change of the resonance field at the phase
transition is discontinuous. In conventional molecular field
approximation, the shift of the antiferromagnetic resonance field
with respect to the paramagnetic resonance is proportional to the
magnitude of the order parameter, i.e., to the sublattice
magnetization. The discontinuous change of the resonance field at
the phase transition indicates that the order parameter is not small
even just below the transition temperature. This observation is in
agreement with the first-order nature of the phase transition in
\zncr{}.\cite{lee:00, plumier-firstorder}

The magnetic susceptibility of the paramagnetic phase is isotropic
due to its cubic symmetry. Cubic symmetry is lost in the ordered
state because of the lattice deformation. The lattice-strain
direction can take one of the equivalent crystallographic axes.
The susceptibility tensor of the antiferromagnet is anisotropic,
the orientation of its principal axes is determined by the orientation
of the order parameter fixed by anisotropic interactions
with respect to the crystallographic axes. Therefore, the
susceptibility tensors of different structural domains are
oriented differently and the gain in the Zeeman energy is
different for different domains. Thus, the application of
magnetic field makes one of the domains more favorable. It
provides an obvious mechanism for the observed field-cooling
effect and for the instability of the resonance absorption close
to $T_N$, described in the previous section. This assumption is in
agreement with the increase of the magnetic susceptibility in the
field-cooled sample (Figure \ref{fig:m(h)}).

The formation of the domain structure in the case of a
cubic-to-tetragonal lattice transition is well studied for
ferroelastic systems. A complicated domain structure consisting of
thin twinned domains is usually formed in ferroelastics (see,
for example, recent Refs.~\onlinecite{jacobs,salje}). Twinning
allows to avoid strong local strain at the contact of the domains
with different directions of deformation axes. The thickness of
twin domains observed in the doped ferroelastic HTSC compound
YBa$_2$Cu$_3$O$_7$ is about 10---100 nm.\cite{salje}  Twinning
also leads to a slight tilting (of the order of $\Delta a/a$) of
the domain axes from the corresponding crystallographic axes
of the high-temperature phase. While the axes tilting is too small
($\Delta a/a\sim 10^{-3}$) to be of importance, small domain
thickness could result in excitation of standing
spin-waves with $k\sim 1/L$ ($L$ is a domain thickness) instead of
a uniform $k=0$ oscillation. This size effect can be a possible
reason for the slight shift of the resonance absorption position
in the FC sample. This allows to estimate the thickness of the
crystallographic domains. We assume a quadratic spectrum of
antiferromagnetic spin-waves

\begin{equation}\label{eqn:size-effect-estimation}
  E=\sqrt{\Delta_0^2+\alpha^2k^2}
\end{equation}

\noindent here $\alpha\sim Ja$. $J$ denotes the exchange coupling constant
between nearest-neighbour spins at distance $a$. If
$\alpha k\ll\Delta_0$, the effective zero-field gap for the
standing spin-waves is larger than for the uniform
oscillation by $\delta=\alpha^2k^2/(2\Delta_0)$. Consequently,
the resonance branches of the monodomain (FC) sample  should be shifted
downwards (on the \textit{H-f} plane) by $\delta$ with respect to the
resonance branches of the multidomain (ZFC) sample: i.e. for the
branches rising with the field the remaining absorption component
of the FC sample shifts at the given frequency  to  higher
fields with respect to its position in the case of the ZFC sample.
This slight shift is observed in the experiment (Figure
\ref{fig:fczfcall}), its magnitude does not exceed the half-width of
the absorption line. We estimate $\delta/h$ as 1 GHz ($h$ is the
Planck constant). Then, the domain thickness can be estimated as

\begin{equation}\label{eqn:thickness}
  \frac{L}{a}\sim\frac{J}{\sqrt{\Delta_0\delta}}
\end{equation}

\noindent which yields (substituting $J=20$~K,\cite{garcia}
$\Delta_0/h$=20~GHz, ) $L/a\sim 100\gg1$.

In further discussion we assume that the domains are thick enough
to be considered as bulk antiferromagnet, and that the domain
walls do not contribute to the magnetic resonance absorption.

\subsection{Application of the exchange-symmetry theory}

We use the theory of exchange symmetry\cite{AndMar} to analyze the
experimental results. According to this theory, if the
relativistic effects are much smaller than the exchange
interaction, the order parameter can be represented by at most 3
unitary orthogonal vectors which transform by irreducible
representations of the crystal symmetry group. The number of
vectors and  these representations define the exchange symmetry of
the magnet. This approach allows to describe all symmetry-based
properties of a magnet without considering its detailed
microscopic structure or any model assumption.  For the case of a
noncollinear antiferromagnet the order parameter consists of at
least two vectors. In case of only two vectors for the sake of
simplicity we denote $\lvec{3}=[\lvec{1}\times\lvec{2}]$.

We derive the dynamic equations for homogeneous oscillations from
the Lagrange function
\begin{equation}\label{eqn:Lagrangian}
{\cal{L}}=\sum_i{\frac{I_i}{2}\left(\dot\lvec{i}+
\gamma[\lvec{i}\times\mathbf{H}]\right)^2}-U_a,
\end{equation}
\noindent where $\gamma$ is the gyromagnetic ratio of the free
electron, the constants $I_i$  are related to the eigenvalues of the
susceptibility tensor, and  the term $U_a$ includes small
relativistic corrections to the main exchange part due to
spin-orbital and dipole-dipole effects. These corrections can be
expanded by the components of the order parameter. The first order
of the $U_a$ expansion is quadratic on $\lij{i}{j}$. The second
order terms can be omitted, because they are $\alpha^2$ times
smaller, where $\alpha$ is the fine-structure constant. The Lagrange
function must be invariant under all transformations of the
crystallographic symmetry group, which results in some relations
between the coefficients in $U_a$ expansion. These relations vary
for different exchange-symmetry groups.

The dynamic equations are obtained by taking a variational
derivative of the Lagrangian (\ref{eqn:Lagrangian}) over the
rotations of the spin space. These equations should be linearized
near the equilibrium position to obtain the eigenfrequencies of
small oscillations. In general case, if the magnetic field direction
deviates from the  eigenvector of the susceptibility tensor, the
equilibrium positions cannot be determined analytically. To model
the observed frequency-field dependences, we perform numerical
calculations of the oscillation eigenfrequencies. We use standard
minimization routines to find an equilibrium orientation of the
order parameter. This modeling procedure is combined with a fitting
algorithm using  the constants $I_i$ and the coefficients of the
$U_a$ expansion as fit parameters.

When performing the expansion of the relativistic corrections, it is
necessary to take into account that in the case of \zncr{} the
magnetic unit cell is larger than the crystallographic one
\cite{lee-jpcm,oles-structure}. Therefore some components of the
order parameter are not invariant under some of the translational
elements of the crystallographic symmetry group. Since the
crystal-symmetry group $D_{2d}^9$, suggested in Ref.
\onlinecite{lee-jpcm}, has a point symmetry $D_{2d}$ in the vertex
of the crystallographic cell, we will focus primarily on the
point-symmetry subgroup. Note that this special property remains for
all subgroups of $D_{2d}^9$. Thus, on discussing the lowering of the
lattice symmetry below $T_N$ to $D_2^7$, we will focus primarily on
the point-symmetry subgroup.

Although some representations of $D_{2d}$ allow weak ferromagnetism,
the susceptibility measurements do not reveal any spontaneous
magnetization. This can be either due to the spontaneous
magnetization being too small, or, more likely, there is no weak
ferromagnetism for the exchange group in the present case.
Therefore, we will not take weak ferromagnetism into account in
further discussion.

\subsection{Evidence for orthorhombic distortions below $T_N$}

Here we will demonstrate that the assumption of the tetragonal
lattice symmetry in the ordered phase contradicts the experimental
observation described in the previous section and the explanation of
the experimental findings requires a further reduction to
orthorhombic symmetry.

First, we note that the symmetry of the magnetic structure below $T_c$ is
lower than tetragonal. This statement follows directly from the
observation of the distinct field-cooling effect for
$\mathbf{H}||\langle111\rangle$, since this field orientation is
equivalent for all tetragonal domains.

For the tetragonal lattice deformation only three types of
different crystallographic domains, differing by the direction of
the tetragonal axis $z$ ($z||\langle001\rangle,\langle010\rangle,
\langle001\rangle$), can be formed at the transition. Since
the magnetic symmetry of the ordered phase
is lower than the lattice symmetry, two types of magnetic domains
with different $x$ and $y$ axes can be formed in each
crystallographic domain. Here and further on in this paper we will
denote by $x$ and $y$ the directions of the two-fold axes
perpendicular to the $z$ axis ($S_4$ axis for the $D_{2d}$
symmetry). Note that the reduction of the point symmetry from
$O_h$ to $D_{2d}$ could be done in two ways: (i) by removal of the
$[100]$ and $[010]$ symmetry axes as well as $(110)$ and
$(1-10)$ mirror planes; (ii) by removal of the $[110]$ and
$[1-10]$ symmetry axes as well as  $(100)$ and $(010)$
mirror planes with the former four-fold axes $\langle100\rangle$
and $\langle010\rangle$ becoming two-fold axes. In the second choice
of axes the magnetic field aligned along the $\langle111\rangle$ direction
would be equivalent for all domains and there would be no reason for the
observed field-cooling effects. Thus, the first possibility has to
be realized. (Identification\cite{lee-jpcm} of the tetragonal
phase as $I\overline{4}m2$ also points to the first possibility.)
Therefore, $x$ and $y$ axes are aligned along the diagonals of the
cubic facets.

\begin{table}[bp]
\caption{Classification of the AFM domains with respect to the
cubic axes of the paramagnetic phase}\label{table}
\begin{center}
\begin{tabular}{|c|c|c|c|}
    \hline
    domain  &$x$  &$y$      &$z$    \\
    \hline
    (a)     &[110]  &[-110] &[001]\\
    (b)     &[1-10] &[110]  &[001]\\
    (c)     &[101]  &[10-1] &[010]\\
    (d)     &[-101] &[101]  &[010]\\
    (e)     &[011]  &[0-11] &[100]\\
    (f)     &[01-1] &[011]  &[100]\\
    \hline

\end{tabular}
\end{center}
\end{table}

Now all domains can be  classified by the orientation of their
"$xyz$" basis with respect to the cubic axes of the paramagnetic
phase as shown in  Table \ref{table}. Some of these domains appear
to be equivalent in the particular experimental conditions (here
$\mathbf{x,y,z}$ are the unit vectors in the corresponding
directions, trivial cases are combined):

\noindent$\mathbf{H}||[100]$
\begin{itemize}
  \item[]domains (e),(f): $\mathbf{H}||\mathbf{z}$
  \item[]domains (a), (b), (c), (d): $\mathbf{H}||\mathbf{x+y}$
\end{itemize}

\noindent$\mathbf{H}||[110]$
\begin{itemize}
  \item[]domain (a): $\mathbf{H}||\mathbf{x}$
  \item[]domain (b): $\mathbf{H}||\mathbf{y}$
  \item[]domains (c),(f), (d), (e):
  $\mathbf{H}||\frac{\mathbf{x+y}}{\sqrt{2}}+\mathbf{z}$
\end{itemize}

\noindent$\mathbf{H}||[111]$
\begin{itemize}
  \item[]domains (a),(c),(e): $\mathbf{H}||\sqrt{2}\mathbf{x}+\mathbf{z}$
  \item[]domains (b),(d),(f): $\mathbf{H}||\sqrt{2}\mathbf{y}+\mathbf{z}$
\end{itemize}

Note that for the $\mathbf{H}||\langle111\rangle$ field
orientation there are only two types of different magnetic
domains. As one can see from the experimental data in Figure
\ref{fig:fit3} we observe five resonance branches in this
orientation: two originating from the higher gap, two originating
from the lower gap and one in the high-field---low-frequency
part of the frequency-field diagram. The observation of this fifth
branch indicates the existence of a third zero-field gap which is
not observed directly and is less or equal in magnitude than the
lowest observed gap of 21~GHz.

To apply the exchange-symmetry theory as described in the previous
subsection it is necessary to specify the symmetry of the order parameter.
The point symmetry group $D_{2d}$ has two-dimensional and
one-dimensional irreducible representations. If $\mathbf{l}^{(1)}$
and $\mathbf{l}^{(2)}$ transform by a two-dimensional
representation, then $I_1=I_2$ and the relativistic contribution
to the Lagrangian (\ref{eqn:Lagrangian}) has the form

\begin{eqnarray}
  U_a&=&\frac{1}{2}A\left(\left(\lx{1}\right)^2+\left(\ly{2}\right)^2\right)+B\left(\lx{1}\ly{2}+\ly{1}\lx{2}\right)+\nonumber\\
  &&+C\left(\lx{1}\ly{2}-\ly{1}\lx{2}\right)+\frac{1}{2}D\left(\lz{3}\right)^2
  \label{eqn:Ua-D2d-1}
\end{eqnarray}

If $\lvec{1}$ and $\lvec{2}$ transform differently under
translations, then $B=C=0$. In the other fundamental case, if
$\lvec{1}$ and $\lvec{2}$ transform by one-dimensional
representations, the relativistic contribution reads

\begin{equation}\label{eqn:Ua-D2d-2}
  U_a=\frac{1}{2}A\left(\lz{1}\right)^2+\frac{1}{2}B\left(\lz{2}\right)^2+C\left(\lx{1}\ly{2}-\ly{1}\lx{2}\right)
\end{equation}

\noindent or

\begin{equation}\label{eqn:Ua-D2d-3}
  U_a=\frac{1}{2}A\left(\lz{1}\right)^2+\frac{1}{2}B\left(\lz{2}\right)^2+C\left(\lx{1}\ly{2}+\ly{1}\lx{2}\right)
\end{equation}

Depending on the coefficients of the $U_a$ expansion,
there are two general possibilities: (i) the $\lvec{i}$ lie in the
$(xz)$ and $(yz)$ planes; (ii) the $\lvec{i}$ lie in the planes
including the $z$ axis and the bisector of the $(xy)$ plane (i.e.
in the $D_{2d}$ group mirror planes). \zncr{} corresponds to the
first case, since in the second case for
$\mathbf{H}||\langle111\rangle$ the order parameter vectors
$\lvec{1}$, $\lvec{2}$, $\lvec{3}$ in all magnetic domains form
the same angles with the field direction, which should result in
the absence of the field cooling effects.

We tried to fit the experimentally observed frequency-field
dependences  applying the exchange-symmetry theory as described
above. However, we did not reach any reasonable agreement of the
modeled curves with the experiment. The reason of this
disagreement can be explained qualitatively. Namely, it can be
proven (see Appendix) that for the tetragonal $D_{2d}$ point
symmetry the domain manifesting a spin reorientation transition
becomes indistinguishable from the other domains above the
transition field. This is due to the fact that even for a magnetic symmetry
lower than tetragonal, the anisotropic contribution $U_a$ reflects the
tetragonal crystal symmetry. Thus, after the spin reorientation transition the
reoriented domain appears to be in a state with the Zeeman
energy lower than before reorientation, but with the same
value of the  anisotropic term $U_a$. The equivalence of the magnetic
domains above the transition should provide the same magnetic
resonance frequencies above the spin-flop transition. Moreover,
there would be no reason for these domains to split again as the
magnetic field is reduced to zero. This, however, contradicts the
experimental observation of the specific resonance modes
corresponding to the domain undergoing a spin-reorientation above
the spin-flop field, as well as to the reproducibility of the
low-field domain structure.

The problem can be solved by the assumption that the
lattice deformation at $T_c$ involves not only a compression along
the $c$-direction, but also a weak in-plane deformation leading to
further reduction of the symmetry. There are two options: to
exclude mirror planes (which yields $D_2$ point symmetry) or to
exclude second-order axes ($C_{2v}$). As we have shown above, the
components of the order parameter lie in the $(xz)$ and $(yz)$
planes. This limits our choice to $D_2$ point symmetry (the
corresponding space subgroup of $D_{2d}^9$ is $D_2^7$ ($F222$))
with second-order axes along $\langle001\rangle$ directions ($z$),
and along $\langle110\rangle$ directions ($x$ and $y$). This
assumption provides inequivalent crystallographic domains
differing by the $x$ and $y$ directions. These orthorhombic
distortions were not detected in the earlier structural studies
\cite{lee:00, lee-jpcm}, most likely because they are probably smaller than
the experimental resolution.

\subsection{Modeling of the AFM resonance frequency-field dependences for
the case of orthorhombic distortions}

We modeled the antiferromagnetic resonance frequency-field
dependences assuming orthorhombic distortions below $T_N$. As was
explained above, to write down the expansion of $U_a$ we again
will focus on the point-symmetry group. The $D_2$ symmetry group
has four one-dimensional representations. There are only four
fundamentally different cases, the others can be reduced to them
by renaming the axes $x,$ $y$ or vectors $\lvec{1}$, $\lvec{2}$:

\begin{eqnarray}
  U_{a}&=&\frac{A}{2}\left(\lz{1}\right)^2+\frac{B}{2}\left(\lz{2}\right)^2+\frac{C}{2}\left(\lx{1}\right)^2+\frac{D}{2}\left(\lx{2}\right)^2+\nonumber\\
  &&+E\left(\lx{1}\ly{2}-\ly{1}\lx{2}\right)+F\left(\lx{1}\ly{2}+\ly{1}\lx{2}\right)\label{eqn:anis1}
\end{eqnarray}

\begin{eqnarray}
  U_{a}&=&\frac{A}{2}\left(\lz{1}\right)^2+\frac{B}{2}\left(\lz{2}\right)^2+\frac{C}{2}\left(\lx{1}\right)^2+\frac{D}{2}\left(\lx{2}\right)^2+\nonumber\\
  &&+E\left(\ly{1}\lz{2}-\lz{1}\ly{2}\right)+F\left(\ly{1}\lz{2}+\lz{1}\ly{2}\right)\label{eqn:anis2}
\end{eqnarray}

\begin{eqnarray}
  U_{a}&=&\frac{A}{2}\left(\lz{1}\right)^2+\frac{B}{2}\left(\lz{2}\right)^2+\frac{C}{2}\left(\lx{1}\right)^2+\frac{D}{2}\left(\lx{2}\right)^2+\nonumber\\
  &&+E\lz{1}\lz{2}+F\lx{1}\lx{2}\label{eqn:anis4}
\end{eqnarray}

\begin{equation}\label{eqn:anis3}
U_{a}=\frac{A}{2}\left(\lz{1}\right)^2+\frac{B}{2}\left(\lz{2}\right)^2+\frac{C}{2}\left(\lx{1}\right)^2+\frac{D}{2}\left(\lx{2}\right)^2
\end{equation}

\noindent Here we again exclude weak ferromagnetism from the
consideration. The first three cases are feasible only if
$\lvec{1}$ and $\lvec{2}$ transform in the same way under
translations. The third case is feasible if $\lvec{1}$ and
$\lvec{2}$ transform by the same one-dimensional irreducible
representation.

The low symmetry of the ordered state results in too many free
parameters in the equations of spin dynamics (four to six
coefficients in the $U_a$ expansion and two of the three $I_i$
constants). By fixing the zero-field gaps of the AFM resonance
spectrum we can put only three constraints on these parameters.
Other constraints are expected to appear during the fitting of the
modeled AFM resonance spectra. This involves too many degrees of
freedom for the assumptions on the sort of equilibrium position,
the way the spin-flop transition takes, and the correspondence
between structural domains and resonance branches. An unequivocal
quantitative reproduction of the experimental curves was difficult
to achieve, however, a reasonable agreement with the
experimentally observed AFM resonance spectra was obtained for
different sets of parameters and for the different forms of $U_a$.

\begin{figure}
  \centering
  \epsfig{file=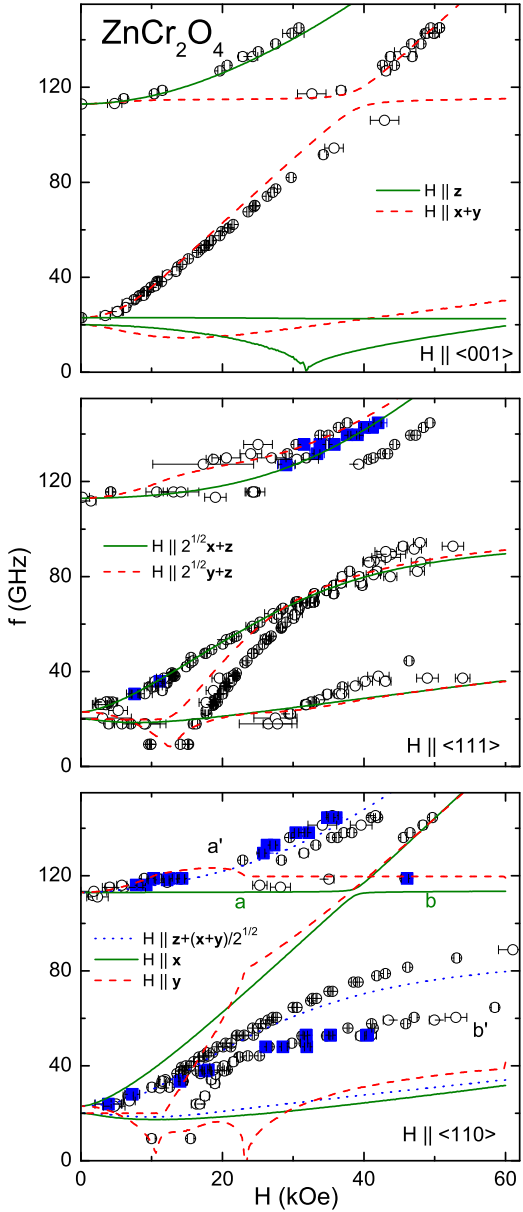, width=\figwidth, clip=}
  \caption{(color online) Frequency-field dependences of the resonance
  modes for the principal cubic orientations (open symbols--ZFC, closed symbols--FC).
  The numeric calculations (lines) are based on Eqn.(\ref{eqn:anis2}).
  Fit parameters: $\gamma=2.8$~GHz/kOe, $I_1$=3.931, $I_2$=0.950,
$I_3$=1.00, $A=65523.87$~GHz$^2$, $B=3444.56$~GHz$^2$,
$C=-6192.39$~GHz$^2$, $D=-5664.49$~GHz$^2$, $E=-8710.32$~GHz$^2$,
$F=0$.
  }\label{fig:fit3}
\end{figure}

Here we present the results of modeling the case of $U_a$ taken in
the form (\ref{eqn:anis2}). From Figure \ref{fig:fit3} one can see
that the correspondence of model and experiment is fairly good.
The correct quantity of resonance branches is obtained. The values
of the zero-field gaps are in good agreement with the experimental
data. For orientations $\mathbf{H}||\langle110\rangle$ and
$\langle111\rangle$ the spin-reorientation transitions of certain
domains are well reproduced.

\begin{figure}
 \epsfig{file=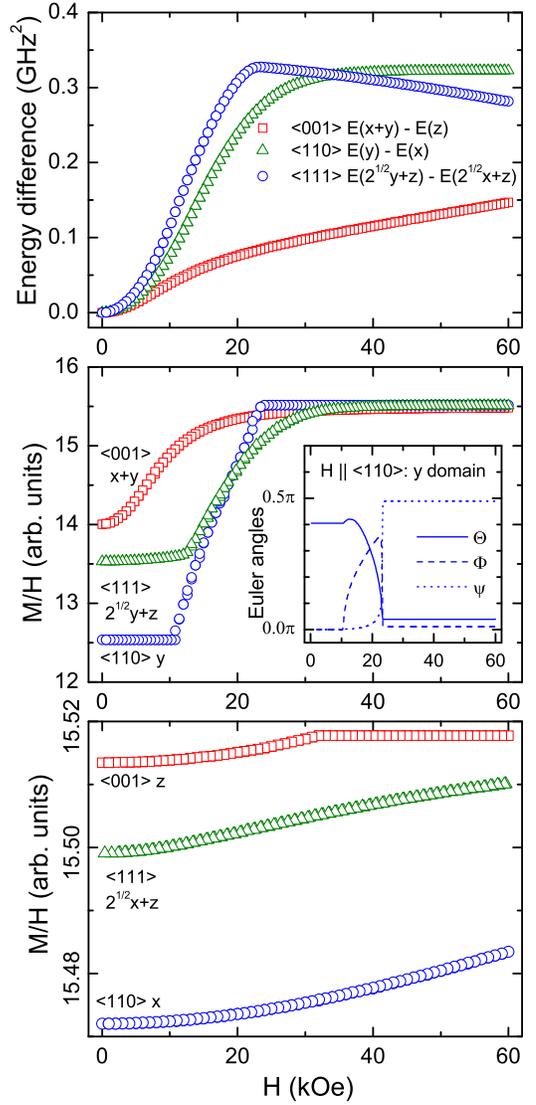, width=\figwidth, clip= }
  \caption{(color online). Upper frame: Calculated field
  dependences of the energy difference for various domains. The
  calculations are done for the same parameters as used for the
  AFM resonance spectrum shown in Figure \ref{fig:fit3}. Middle
  frame: Calculated magnetization of the domains with unfavorable
  orientation. Inset: Calculated field dependences of the Euler
  angles for the $\mathbf{H}||\langle110\rangle$ $y$-domain. Lower
  frame: Calculated magnetization of the domains with favorable orientation
  of the order parameter.}\label{fig:e_sub}
\end{figure}

Our modeling indicates that the frequency-field dependences can be
reasonably reproduced assuming a single type of order parameter
and taking into account all possible domains. There are really
just two types of discrepancies between the experimental data and
modeled curves: first, there are predicted low-frequency modes at
high fields which are not observed; second, for the orientation
$\mathbf{H}||\langle110\rangle$ the modes corresponding to the
domain with lowest energy ($\mathbf{H}||x$) are far from the
experimental data corresponding to the domain stable under field
cooling (see the fit curves and experimental data marked by a, a',
b, b' in the lower panel of Fig.~\ref{fig:fit3}). These
discrepancies can be due to the following reasons: at 9~GHz ESR
experiments, fields up to only 18~kOe were available,  whereas
above 18~GHz the frequency-field dependences of the low-frequency
modes in question are rather flat. In the case of $I_2=I_3$ they
become field independent which results in a strong broadening of
the resonance modes in the experiments at fixed frequency. Thus,
these modes cannot be detected in a field scan. Concerning the
discrepancy for the $\mathbf{H}||x$ domain, note that the modeled
curves still demonstrate an "anti-crossing" which can be enhanced,
for example, by higher order terms in the energy expansion.

Our modelling demonstrates that the spin-reorientation transition can be realized by a
continuous rotation of the order parameter as shown e.g.
for the $y$ domain in the inset of Fig.~\ref{fig:e_sub}.
This explains the absence of a sharp change of the magnetization
at the spin-reorientation transition which is simulated in the
middle and lower frames of this figure. The field dependence of the magnetization is
qualitatively well reproduced: it demonstrates a nonlinear behavior
with a characteristic change during the spin-reorientation for the
unfavorable domains, which would disappear after field cooling,
and an almost linear behavior for the favorable domains which
remains after field cooling. The energy difference between different
domains shown in the upper frame of this figure explains
the observed field-cooling phenomena. The domains corresponding to
the modes surviving the field cooling have the lowest energy in
the magnetic field. For the orientation $\mathbf{H}||\langle111\rangle$
the calculated energy difference
between the domains is similar and, thus, the field-cooling effect
is comparable to the case $\mathbf{H} || \langle110\rangle$.
For $\mathbf{H}||\langle001\rangle$ this difference is
much smaller which can probably explain the much weaker
field-cooling effect for this direction.
However we have to admit the problem in reproducing the experimentally observed
anisotropy of the field-cooled magnetization at low fields: the experimentally highest
susceptibility for the monodomain FC sample is observed for
$\mathbf{H}||\langle110\rangle$ (Figure \ref{fig:m(h)}), while the
modelled curves for the presented parameter set indicate a
slightly higher susceptibility for the most favorable domain in
the $\mathbf{H}||\langle100\rangle$ (Figure \ref{fig:e_sub}). Note that the modeled difference
of the three favorable domains is tiny, thus for low fields some weak effects related to shape anisotropy
or surface effect may distroy the stability of the uniform magnetization.

Summarizing the results of our modeling, we conclude that the
low-energy dynamics of the ordered phase of \zncr{} can be described
by the assumption of a single type of magnetic order below $T_N$.
This puts a question mark on the possibility to realize different
spin structures in \zncr{} as suggested earlier \cite{chung-prl}. We
suppose that the complications in the determination of the magnetic
structure of \zncr{} by neutron scattering were caused by the
unaccounted orthorhombic deformations and effects of multiple
domains.

\section{Conclusions}

We have performed a detailed study of the low-energy dynamics of the
ordered phase of the frustrated  antiferromagnetic spinel \zncr{}.
We have proven directly that multiple domains exist below the
transition temperature $T_N$. We have demonstrated  that some of
these domains are effectively suppressed by field cooling.
Spin-reorientation transitions indicated by softening of certain
antiferromagnetic resonance modes and by the nonlinear behavior of
the magnetization are observed.

These results are incompatible with the earlier proposed symmetry of
the distorted lattice. Thus, we conclude that the lattice
deformation at the phase transition involves small in-plane
distortions besides the tetragonal distortions. We suggest that the
actual symmetry of the lattice below $T_N$ corresponds to the
orthorhombic $D_2^7$ symmetry.

We have demonstrated that the low-energy dynamics can be reasonably
described within the framework of the exchange-symmetry theory,
assuming a noncollinear magnetic ordering characterized by a single
order parameter.

\acknowledgements

Authors thank A.I.~Smirnov, S.S.~Sosin, V.I.~Marchenko,
L.E.~Svistov, I.N.~Khliustikov for the continuous interest and
numerous discussions. The work was supported by the RFBR grant
No.07-02-00725, by the Russian Science Support Foundation, Russian
Presidential Grant for the Young Scientists MK-4569.2008.2,
Russian Presidential Grants for the Support of Scientific Schools
6122.2008.2 and 3526.2008.2 and partially by the Deutsche
Forschungsgemeinschaft (DFG) within the collaborative research center SFB 484 (Augsburg).

\appendix

\section{Instability of the certain domains in case of the tetragonal lattice symmetry}

Here we will  prove that for the $D_{2d}$ point symmetry at the
spin-reorientation transition the domain demonstrating this
transition becomes indistinguishable from the other domains.

First we consider the case when $\lvec{1}$ and $\lvec{2}$ transform
by two-dimensional representation of $D_{2d}.$ Minimizing the energy
(\ref{eqn:Ua-D2d-1}) we get the orientation of the order parameter
in the absence of magnetic field. For $|B|>|A|$ the solution
consists vectors lying in the mirror planes, which is not the case
of \zncr{} (see above). For $|A|>|B|$ and $A<0$ the two solutions
are $\lvec{1}||\mathbf{x}$, $\lvec{2}$ lying in the $(yz)$ plane and
$\lvec{2}||\mathbf{y},$ $\lvec{1}$ lying in the $(xz)$ plane. The
case of $A>0$ can be reduced to this one by renaming the axes $x$
and $y$. These solutions define two different magnetic domains. If
the susceptibility along $\lvec{3}$ is less than along $\lvec{1}$
and $\lvec{2}$, then for $H||\mathbf{x}$ one of the domains is
already in its minimum of the Zeeman energy. The other one is not in the minimum and at
some value of magnetic field it will undergo a spin-reorientation
transition. As it turns out, there is only one possible state for it
after the transition, the same as for the first domain. So after the
spin-flop these two domains will be indistinguishable. If the
susceptibility along $\lvec{1}$ and $\lvec{2}$ is less than along
$\lvec{3}$, both domains are not in the minimum of the Zeeman energy
when the field is applied along the $x$ axis. Still after the
spin-flop both domains again become indistinguishable.

Now we  consider case when $\lvec{1}$ and $\lvec{2}$  transform by
one-dimensional representations of $D_{2d}$. To define the
orientation  of the order parameter  we minimize the energy
(\ref{eqn:Ua-D2d-2}). We get the result that one of the vectors (let
it be $\lvec{1}$) is aligned in the $(xy)$ plane, but its
orientation in this plane remains arbitrary. To find a solution, it
is necessary to take into account the next order terms in the $U_a$
expansion. There is no need to write down all of them, just note
that due to the tetragonal symmetry the dependence of $U_a$ on the
angle $\phi$ between $\lvec{1}$ and $\mathbf{x}$ is $F\cos(4\phi)$.
There are two sets of solutions depending on the sign of $F.$ For
$F>0$ the solutions are $\phi=\pi/4,3\pi/4,5\pi/4,7\pi/4$, i.e.
$\lvec{1}$ lies in the mirror plane, which is not the case of
\zncr{}. For $F<0$ the solutions are $\phi=0,\pi/2,\pi,3\pi/2$.
These solutions define two magnetic domains: for one of them
$\lvec{1}||\mathbf{x}$ and for the other $\lvec{1}||\mathbf{y}.$
First, no matter along which $\lvec{i}$ the susceptibility is
largest, a spin-reorientation transition is expected for
$\mathbf{H}||(\mathbf{x}\pm\mathbf{y})$ (i.e.
$\mathbf{H}||\langle001\rangle$). This spin-reorientation is either
rotation of the order parameter around the $z$ axis by $\pi/4$ or,
in the special case of the largest susceptibility being along the
$z$ axis, rotation of the largest susceptibility direction to the
$(xy)$ plane. However, such a transition is not observed in our
experiments. Second, the spin-flop transition observed at
$\mathbf{H}||\langle111\rangle$ can be caused only by rotation of
the order parameter around the $z$ axis, but after this rotation both
domains become indistinguishable.


\begin{thebibliography}{10}


\bibitem{huang:04} D. J. Huang, C. F. Chang, H.-T. Jeng, G. Y. Guo,
H.-J. Lin, W. B. Wu, H. C. Ku, A. Fujimori, Y. Takahashi, and C. T.
Chen, Phys. Rev. Lett. \textbf{93}, 077204 (2004).

\bibitem{leonov:04}I. Leonov, A. N. Yaresko, V. N. Antonov, M. A. Korotin,
and V. I. Anisimov, Phys. Rev. Lett. \textbf{93}, 146404
(2004).

\bibitem{kondo:97}S. Kondo, D. C. Johnston, C. A. Swenson,
F. Borsa, A. V. Mahajan, L. L. Miller, T. Gu, A. I. Goldman,
 M. B. Maple, D. A. Gajewski, E. J. Freeman, N. R. Dilley, R. P. Dickey,
 J. Merrin, K. Kojima, G. M. Luke, Y. J. Uemura,
 O. Chmaissem, J. D. Jorgensen,
 Phys. Rev. Lett.
\textbf{78}, 3729 (1997).

\bibitem{krimmel:99}A. Krimmel, A. Loidl, M. Klemm, S. Horn, and H. Schober,
 Phys. Rev. Lett.
\textbf{82}, 2919 (1999).


\bibitem{ramirez:97}A.~P. Ramirez, R.~J. Cava, J. Krajewski, Nature (London)
\textbf{386}, 156 (1997).

\bibitem{fritsch:03} V. Fritsch, J. Deisenhofer, R. Fichtl, J. Hemberger,
H.-A. Krug von Nidda, M. M\"ucksch, M. Nicklas, D. Samusi, J. D.
Thompson, R. Tidecks, V. Tsurkan, and A. Loidl, Phys. Rev. B
\textbf{67}, 144419 (2003).

\bibitem{ogushi:05}K. Ohgushi, T. Ogasawara, Y. Okimoto, S. Miyasaka, and Y.
Tokura, Phys. Rev. B \textbf{72}, 155114 (2005).

\bibitem{fichtl:05}R. Fichtl, V. Tsurkan, P. Lunkenheimer, J. Hemberger,
V. Fritsch, H.-A. Krug von Nidda, E.-W. Scheidt, and A. Loidl, Phys.
Rev. Lett. \textbf{94}, 027601 (2005).

\bibitem{fritsch:04}V. Fritsch, J. Hemberger, N. B\"uttgen, E.-W. Scheidt,
H.-A. Krug von Nidda, A. Loidl, and  V. Tsurkan, Phys. Rev. Lett.
\textbf{92}, 116401 (2004).

\bibitem{hemberger:05}J. Hemberger,  P. Lunkenheimer, R. Fichtl,
H.-A. Krug von Nidda, V. Tsurkan, and A. Loidl, Nature (London)
\textbf{416}, 364 (2005).

\bibitem{weber:06}S. Weber, P. Lunkenheimer, R. Fichtl, J. Hemberger,
 V. Tsurkan, and A. Loidl, Phys. Rev.Lett.
 \textbf{96}, 157202 (2006).

\bibitem{hemberger:07}J. Hemberger, H.-A. Krug von Nidda,  V. Tsurkan,
and  A. Loidl, Phys. Rev. Lett.
\textbf{98}, 147203 (2007).

\bibitem{hemberger:06}J. Hemberger, T. Rudolf, H.-A. Krug von Nidda,
F. Mayr, A. Pimenov,V. Tsurkan, and A. Loidl, Phys. Rev. Lett.
\textbf{97}, 087204 (2006).

\bibitem{rudolf:07}T. Rudolf, C. Kant, F. Mayr, J. Hemberger,
V. Tsurkan, and A. Loidl, Phys. Rev. B
\textbf{75}, 052410 (2007).

\bibitem{radaelli:02}P. G. Radaelli, Y. Horibe, M. J. Gutmann, H. Ishibashi,
C. H. Chen, R. M. Ibberson, Y. Koyama, Y. S. Hor, V. Kiryukhin, S. W. Cheong,
Nature (London) \textbf{416}, 155 (2002).

\bibitem{schmidt:04}M. Schmidt, W. Ratcliff II, P. G. Radaelli,
K. Refson, N. M. Harrison, and S. W. Cheong Phys. Rev. Lett.
\textbf{92}, 056402 (2004).

\bibitem{lee:00} S.-H. Lee, C. Broholm, T.H. Kim, W. Ratcliff, II, and
S-W. Cheong, Phys. Rev. Lett. \textbf{84}, 3718
(2000).

\bibitem{tchernyshyov:02a}O. Tchernyshyov, R. Moessner, and
S. L. Sondhi, Phys. Rev. Lett. \textbf{88}, 067203 (2002).

\bibitem{tchernyshyov:02b}O. Tchernyshyov, R. Moessner, and S. L. Sondhi,
 Phys. Rev. B \textbf{66}, 064403 (2002).


\bibitem{chung-prl} J.-H. Chung, M. Matsuda, S.-H. Lee, K. Kakurai,
H. Ueda, T.J. Sato, H. Takagi, K.-P. Hong, and S. Park, Phys. Rev.
Lett. \textbf{95}, 247204 (2005).

\bibitem{lee-jpcm} S.H. Lee, G. Gasparovich, C. Broholm, M. Matsuda,
J.-H. Chung, Y.J. Kim, H. Ueda, G. Xu, P. Zschak, K. Kakurai, H.
Takagi, W. Ratcliff, II, T.H. Kim and S.W. Cheong, J. Phys: Condens.
Matter \textbf{19}, 145259 (2007).

\bibitem{martinho-prb}H. Martinho, N.O. Moreno, J.A. Sanjurjo,
C. Rettori, A.J. Garc\`{i}a-Adeva, D.L. Huber, S.B. Oseroff, W.
Ratcliff II, S.-W. Cheong, P.G. Pagliuso, J.L. Sarrao, G.B.
Martins, Phys. Rev B \textbf{64}, 024408 (2001).

\bibitem{AndMar} A.F. Andreev, V.I. Marchenko, Sov. Phys. Usp. \textbf{130}, 39
(1980).

\bibitem{plumier-firstorder} R. Plumier, M. Lecomte and M. Sougi,
J. Physique Lett. \textbf{38}, L149 (1977).

\bibitem{jacobs} A.E. Jacobs, S.H. Curnoe and
R.C.Desai, Phys. Rev. B \textbf{68}, 224104 (2003)

\bibitem{salje} E.K.H. Salje, S.A. Hayward and W.T. Lee, Acta
Cryst. Sect. A \textbf{61}, 3 (2005)

\bibitem{garcia} A. J. Garc\'{i}a-Adeva  and D. L. Huber, Phys. Rev. Lett. \textbf{85}, 4598 (2000)

\bibitem{oles-structure}A. Ol\'{e}s, J. Phys. Colloque, \textbf{32}, C1-328
(1971).

\end{thebibliography}
\end{document}